\documentclass[twocolumn,epsfig,showpacs]{revtex4}
\usepackage{epsfig,color}

\begin{document}
\title{Hadronic Matter is Soft\\
}

\author{Ch. Hartnack${}^{1}$, H. Oeschler${}^{2}$ and J\"org Aichelin${}^{1}$}

\address{${}^{1}$SUBATECH,
Laboratoire de Physique Subatomique et des Technologies Associ\'ees \\
University of Nantes - IN2P3/CNRS - Ecole des Mines de Nantes \\
4 rue Alfred Kastler, F-44072 Nantes Cedex 03, France}

\address{${}^{2}$Institut f\"ur Kernphysik,
Darmstadt University of Technology, 64289 Darmstadt, Germany}

\begin{abstract}
The stiffness of the hadronic equation of state has been extracted
from the production rate of $K^+$ mesons in heavy ion collisions
around 1 $A$ GeV incident energy. The data are best described with
a compressibility coefficient $\kappa$ around 200 MeV, a value
which is usually called ``soft''. This is concluded from a
detailed comparison of the results of transport theories with the
experimental data using two different procedures: (i) the energy
dependence of the ratio of $K^+$ from Au+Au and C+C collisions and
(ii) the centrality dependence of the $K^+$ multiplicities. It is
demonstrated that input quantities of these transport theories
which are not precisely known, like the kaon-nucleon potential,
the $\Delta N \rightarrow N K^+ \Lambda$ cross section or the life
time of the $\Delta$ in matter do not modify this conclusion.
\vspace{.6cm}
\end{abstract}

\maketitle

Since many years one of the most important
challenges in nuclear physics is to determine $E/A(\rho,T)$, the
energy/nucleon in nuclear matter in thermal equilibrium as a
function of the density $\rho$ and the temperature $T$. Only at
equilibrium density, $\rho_0$, the energy per nucleon
$E/A(\rho=\rho_0,T=0)= -16$ MeV is known by extrapolation of the
Weizs\"acker mass formula to infinite matter. This quest has been
dubbed ``search for the nuclear equation of state (EoS)''.

Modelling of neutron stars or supernovae have not yet constrained
the nuclear equation of state \cite{SN}. Therefore, the most
promising approach to extract $E/A(\rho,T)$ are heavy ion
reactions in which the density of the colliding nuclei changes
significantly. Three principal experimental observables have been
suggested in the course of this quest which carry - according to
theoretical calculations - information on the nuclear EoS: (i) the
strength distribution of giant isoscalar monopole resonances
\cite{you,mon}, (ii) the in-plane sidewards flow of nucleons in
semi-central heavy ion reaction at energies between 100 $A$ MeV
and 400 $A$ MeV \cite{sto} and (iii) the production of $K^+$
mesons in heavy ion reactions at energies around 1 $A$ GeV
\cite{aik}. Although theory has predicted these effects
qualitatively, a quantitative approach is confronted with two
challenges: a) The nucleus is finite and surface effects are not
negligible, even for the largest nuclei and b) in heavy ion
reactions the reacting system does not come into equilibrium.
Therefore complicated non-equilibrium transport theories have to
be employed and the conclusion on the nuclear equation of state
can only be indirect.

(i) The study of monopole vibrations has been very successful, but
the variation in density is minute. Therefore, giant monopole
resonances are sensitive to the energy which is necessary to
change the density of a cold nucleus close to the equilibrium
point $\rho_0$. According to theory the vibration frequency
depends directly on the force which counteracts to any deviation
from the equilibrium and therefore to the potential energy. The
careful analysis of the isoscalar monopole strength in
non-relativistic \cite{you} and relativistic mean field models has
recently converged \cite{mon} due to a new parametrization of the
relativistic potential. These calculations allow now for the
determination of the compressibility $\kappa = 9 \rho^2
\frac{d^2E/A(\rho,T)}{d^2\rho} |_{\rho=\rho_0}$ which measures the
curvature of $E/A(\rho,T)$ at the equilibrium point. The values
found are around $\kappa = 240$ MeV and therefore close to what
has been dubbed ``soft equation of state" .

(ii) If the overlap zone of projectile and target becomes
considerably compressed in semi-central heavy-ion collisions, an
in-plane flow is created due to the transverse pressure on the
baryons outside of the interaction region with this flow being
proportional to the transverse pressure. In order to obtain a
noticeable compression, the beam energy has to be large as
compared to the Fermi energy of the nucleons inside the nuclei and
hence a beam energy of at least 100 $A$ MeV is necessary.
Compression goes along with excitation and therefore the
compressional energy of excited nuclear matter is encoded in the
in-plane flow. It has recently been demonstrated \cite{ant} that
transport theories do not agree quantitatively yet and therefore
former conclusions \cite{paw2} have to be considered as premature.

(iii) The third method is most promising for the study of nuclear
matter at high densities  and is subject of this Letter. $K^+$
mesons produced far below the $NN$ threshold cannot be created in
first-chance collisions between projectile and target nucleons.
They do not provide sufficient energy even if one includes the
Fermi motion. The effective energy for the production of a $K^+$
meson in the $NN$ center of mass system is 671 MeV as in addition
to the mass of the kaon a nucleon has to be converted into a
$\Lambda$ to conserve strangeness. Before nucleons can create a
$K^+$ at these subthreshold energies, they have to accumulate
energy. The most effective way to do this is the conversion of a
nucleon into a $\Delta$ and to produce in a subsequent collision a
$K^+$ meson via $\Delta N \rightarrow N K^+ \Lambda$. Two effects
link the yield of produced $K^+$ with the density reached in the
collision and the stiffness of the EoS. If less energy is needed
to compress matter (i) more energy is available for the $K^+$
production and (ii) the density which can be reached in these
reactions will be higher. Higher density means a smaller mean free
path and therefore the $\Delta$ will interact more often
increasing the probability to produce a $K^+$ and hence, it has a
lower chance to decay before it interacts. Consequently the $K^+$
yield depends on the compressional energy. At beam energies around
1 $A$ GeV matter becomes highly excited and mesons are formed.
Therefore this process tests highly excited hadronic matter. At
beam energies $> 2~A$ GeV first-chance collisions dominate and
this sensitivity is lost.

In this Letter we would like to report that for the third approach
different transport theories have converged. Two independent
experimental observables, the ratio of the excitation functions of
the $K^+$ production for Au+Au and for C+C~\cite{sturm,fuchs}, and
a new observable, the dependence on the number of participants of
the $K^+$ yield show that nucleons interact with a potential which
corresponds to a compressibility of $\kappa  \le 200$ MeV in
infinite matter in thermal equilibrium. This value extracted for
hadronic matter at densities around 2.5 times the normal nuclear
matter density is very similar to that extracted at normal nuclear
matter density. A key point of this paper is to demonstrate that
the different implementation of yet unsolved physical questions,
like the $N\Delta \rightarrow K^+ \Lambda N$ cross section, the
$KN$ interaction as well as the life time of the nuclear
resonances in the hadronic environment do not affect this
conclusion.

In order to determine the energy which is necessary to compress
infinite nuclear matter in thermal equilibrium by heavy ion
reactions in which no equilibrium is obtained one chooses the
following strategy: The transport theory calculates the time
evolution of the quantal particles described by  Gaussian wave
functions. The time evolution is given by a variational principle
and the equations one obtains for this choice of the wave function
are identical to the classical Hamilton equations where the
classical two-body potential is replaced by the expectation value
of the real part of the Br\"uckner $G$-matrix. For this potential
the potential energy in infinite nuclear matter is calculated. To
determine the nuclear equation of state we average this
(momentum-dependent) two-body potential over the momentum
distribution of a given temperature $T$ and add to it the kinetic
energy. Expressed as a function of the density we obtain the
desired nuclear equation of state $E/A(\rho,T)$. Our two-body
potential has five parameters which are fixed by the binding
energy of infinite nuclear matter at $\rho_0$, the compressibility
$\kappa$ and the optical potential which has been measured in pA
reactions.

Once the parameters are fixed we use the two-body potential with
these parameters in the transport calculation. There is an
infinite number of two-body potentials which give the same
equation of state because the range of the potential does not play
a role in infinite matter. The nuclear surface measured in
electron scattering on nuclei fixes the range, however, quite
well. The uncertainty which remains is of little relevance here
(in contradiction to the calculation of the in-plane flow which is
very sensitive to the exact surface properties of the nuclei and
hence to the range of the potential).

We employ the Isospin Quantum Molecular Dynamics (IQMD) \cite{har}
approach with the following equations of motion:
\begin{equation}\label{hamiltoneq}
\dot{\vec{p}}_i = - \frac{\partial \langle H \rangle}{\partial \vec{r}_i}
\quad {\rm and} \quad
\dot{\vec{r}}_i = \frac{\partial \langle H \rangle}{\partial \vec{p}_i} \, ,
\end{equation}
where the expectation value of the total Hamiltonian reads as
$\langle H \rangle = \langle T \rangle + \langle V \rangle$
with
\begin{eqnarray}
\langle T \rangle &=& \sum_i \frac{p_i^2}{2m_i} \cr
\langle V \rangle &=& \sum_{i} \sum_{j>i}
 \int f_i(\vec{r},\vec{p},t) \,
V^{ij}  f_j(\vec{r}\,',\vec{p}\,',t)\,
d\vec{r}\, d\vec{r}\,'
d\vec{p}\, d\vec{p}\,' \quad.
\end{eqnarray}
and $f_i$ being the Gaussian Wigner density of nucleon $i$. The
baryon-potential consists of the real part of the $G$-Matrix which
is supplemented by the Coulomb interaction between the charged
particles. The former can be further subdivided in a part
containing the contact Skyrme-type interaction only, a
contribution due to a finite range Yukawa-potential, and a
momentum-dependent part
with
\begin{eqnarray}
V^{ij} 
       &=& V^{ij}_{\rm Skyrme} + V^{ij}_{\rm Yuk} + V^{ij}_{\rm mdi} +
           V^{ij}_{\rm Coul} \nonumber \\
       &=& t_1 \delta (\vec{x}_i - \vec{x}_j) +
           t_2 \delta (\vec{x}_i - \vec{x}_j) \rho^{\gamma-1}(\vec{x}_{i}) +
       \nonumber \\
       & &    t_3 \frac{\hbox{exp}\{-|\vec{x}_i-\vec{x}_j|/\mu\}}
               {|\vec{x}_i-\vec{x}_j|/\mu} + \label{vijdef}
           \frac{Z_i Z_j e^2}{|\vec{x}_i-\vec{x}_j|} +\nonumber  \\
       & & t_4\hbox{ln}^2 (1+t_5(\vec{p}_i-\vec{p}_j)^2)
               \delta (\vec{x}_i -\vec{x}_j)
\end{eqnarray}
with $Z_i,Z_j$ the charges of the baryons $i$ and $j$. For more
details we refer to Ref.~\cite{har}.

We include in this calculation all inelastic cross sections which
are relevant for the $K^+$ production. For details of these cross
sections we refer to \cite{crosss}. Unless specified differently,
the change of the $K^+$ mass due to the kaon-nucleon ($KN$)
interaction according to $m^K(\rho) = m^K_0(1-0.075
\frac{\rho}{\rho_0})$ is taken into account, in agreement with
recent self-consistent calculations of the spectral function of
the $K^+$ \cite{lu}. The $\Lambda$ potential is 2/3 of the nucleon
potential, assuming that the s quark is inert.

In order to minimize the experimental systematical errors and the
consequences of theoretical uncertainties it is better to compare
ratios of cross sections rather than the absolute values
\cite{sturm}. We have made sure that the standard version of IQMD
reproduces the excitation function for Au+Au as well as for C+C
quite well \cite{habil}. These ratios are quite sensitive to the
nuclear potentials because the compression obtained in the Au+Au
collisions is considerable (up to 3$\rho_0$) and depends on the
nuclear equation of state whereas in C+C collisions the
compression is small and almost independent on the stiffness of
the EoS.

\begin{figure}
\vspace*{-.5cm} 
\epsfig{file=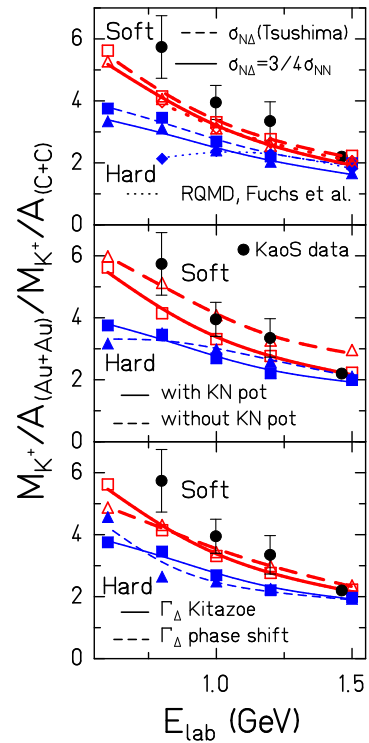,width=7.0cm}
 \caption{Comparison of the measured excitation
function of the ratio of the $K^+$ multiplicities per mass number
$A$ obtained in Au+Au and in C+C reactions (Ref.~\cite{sturm})
with various calculations. The use of a hard EoS is denoted by
thin (blue) lines, a soft EoS by thick (red) lines. The calculated
energies are given by the symbols, the lines are drawn to guide
the eye. On top, two different versions of the $N\Delta
\rightarrow K^+\Lambda N$ cross sections are used. One is based on
isospin arguments \cite{ko}, the other is determined by a
relativistic tree level calculation \cite{tsus}. The calculation
by Fuchs \cite{fuchs} are shown as dotted lines. Middle: IQMD
calculations with and without $KN$ potential are compared. Bottom:
The influence of different options for the life time of  $\Delta$
in matter is demonstrated.} \label{ratio}
\end{figure}

Figure \ref{ratio} shows the comparison of the measured ratio of
the $K^+$ multiplicities obtained in Au+Au and C+C reactions
\cite{sturm} together with transport model calculations as a
function of the beam energy.

We see clearly that the form of the yield ratio depends on the
potential parameters (hard EoS: $\kappa$ = 380 MeV, thin lines and
solid symbols, soft EoS: $\kappa$ = 200 MeV , thick lines and open
symbols) in a quite sensible way and that the prediction in the
standard version of the simulation (squares) for a soft and a hard
EoS potential differ much more than the experimental
uncertainties. The calculation of Fuchs et al.~\cite{fuchs} given
in the same graph, agrees well with our findings.

This observation is, however, not sufficient to determine the
potential parameters uniquely because in these transport theories
several not precisely known processes are encoded. Therefore, it
is necessary to verify that these uncertainties do not render this
conclusion premature.  Figure \ref{ratio}, top, shows as well the
influence of the unknown $N\Delta \rightarrow K^+ \Lambda N$ cross
section on this ratio. We confront the standard IQMD option (with
cross sections for $\Delta N$ interactions from Tsushima et
al.~\cite{crosss}) with another option, $\sigma(N\Delta) = 3/4
\sigma(NN)$~\cite{ko}, which is based on isospin arguments and has
been frequently employed. Both cross sections differ by up to a
factor of ten and change significantly the absolute yield of $K^+$
in heavy ion reactions but do not change the shape of the ratio.

The middle part demonstrates the influence of the kaon-nucleon
potential which is not precisely known at the densities obtained
in this reaction. The uncertainties due to the $\Delta$ life time
are discussed in the bottom part. Both calculations represent the
two extreme values for this lifetime \cite{crosss} which is
important because the disintegration of the $\Delta$ resonance
competes with the $K^+$ production.

Thus we see that these uncertainties do not influence the
conclusion that the excitation function of the ratio is quite
different for a soft EoS potential as compared to a hard one and that
the data of the KaoS collaboration are only compatible with the
soft EoS potential. The only possibility to change this
conclusions is the assumption that the cross sections are
explicitly density dependent in a way that the increasing density
is compensated by a decreasing cross section. 
It would have a strong influence on other observables which are
presently well predicted by the IQMD calculations.

We would like to add that the smoothness of the excitation
function also demonstrates that there are no density isomers in
the density regions which are obtained in these reactions because
the $K^+$ excitation function would be very sensitive to such an
isomeric state \cite{jha}.

\begin{figure}
\vspace*{-.5cm} 
\epsfig{file=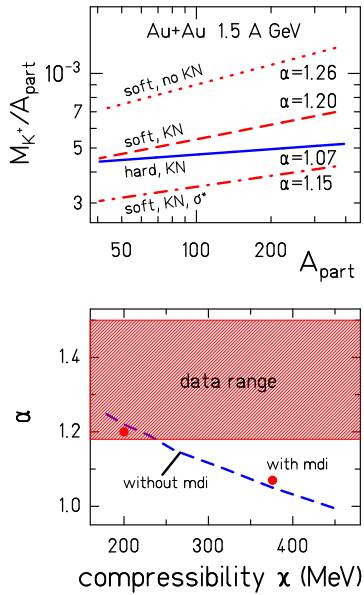,width=5.9cm}
\caption{Dependence of the $K^+$ scaling on the nuclear equation
of state.  We present this dependence in form of $M_{K^+}=A_{{\rm
part}}^\alpha$. On the top the dependence of $M_{K^+}/A_{{\rm
part}}$ as a function of $ A_{{\rm part}}$ is shown for different
options: a ``hard'' EoS with $KN$ potential (solid line), the
other three lines show a ``soft'' EoS, without $KN$ potential and
$\sigma(N\Delta)$ from Tsushima~\cite{tsus} (dotted line), with
$KN$ potential and the same parametrization of the cross section
(dashed line) and with $KN$ potential and $\sigma(N\Delta) = 3/4
\sigma(NN)$. On the bottom the fit exponent $\alpha$ is shown as a
function of the compressibility for calculations with
momentum-dependent interactions (mdi) and for static interactions
($t_4=0$, dashed line).} \label{part}
\end{figure}

The conclusion that nuclear matter is best described by a soft
EoS, is supported by another variable, the dependence of the $K^+$
yield on the number of participating nucleons ${A_{{\rm part}}}$.
The prediction of the IQMD simulations in the standard version for
this observable is shown in Fig.~\ref{part}. The top of the figure
shows the kaon yield $M_{K^+}/A_{{\rm part}}$ for Au+Au collisions
at 1.5 $A$ GeV as a function of the participant number ${A_{{\rm
part}}}$ for a soft EoS using different options: standard version
(soft, $KN$), calculations without kaon-nucleon interaction (soft,
no $KN$) and with the isospin based $N\Delta \rightarrow N\Lambda
K^+$ cross section (soft, $KN$, $\sigma^*$). These calculations
are confronted with a standard calculation using the hard EoS
potential. The scaling of the kaon yield with the participant
number  can be parameterized by $M_{K^+}=A_{{\rm part}}^\alpha$.

All calculations with a soft EoS show a rather similar value of
$\alpha$ - although the yields are very different - while the
calculation using a hard equation shows a much smaller value.
Therefore we can conclude that also the slope value $\alpha$ is a
rather robust observable.

The bottom of Fig.~\ref{part} shows that $\alpha$ depends smoothly
on the compressibility  $\kappa$ of the EoS. Whether we include
the momentum dependence of the nucleon nucleon interaction (with
mdi) or not (without mdi) does not change the value of $\alpha$ as
long as the compressibility is not changed - in stark contrast to
the in-plane flow. Again, the measured centrality dependence for
Au+Au at 1.5 $A$ GeV from the KaoS collaboration \cite{foer},
$\alpha = 1.34 \pm 0.16$, is only compatible with a soft EoS
potential.

This finding is also supported by a more recent analysis
\cite{paw,and} of the in-plane flow which supersedes the former
conclusion that the EoS is hard \cite{hard} (made before the
momentum-dependent interaction has been included in the
calculations). Due to the strong dependence of the in-plane flow
on the potential range parameter and its dependence on the
particles observed these conclusions are much less firm presently.
Comparisons of the out-of-plane squeeze of baryons also show a
preference for a soft equation of state with momentum dependent
interactions\cite{hsqueeze}.

In conclusion, we have shown that the two experimental observables
which are most sensitive to the potential parameters of the
nucleon-nucleon interaction are only compatible with those
parameters which lead in nuclear matter to a soft hadronic EoS.
This conclusion is robust. Uncertainties of the input in these
calculations, like the $KN$ potential at high densities, the
lifetime of the $\Delta$ in matter and the $\Delta N \rightarrow N
K^+ \Lambda$ cross section do not influence this conclusion. The
potential parameter $\kappa$ is even smaller than that extracted
from the giant monopole vibrations. Thus the energy which is
needed to compress hadronic matter of $\kappa \le 200$ MeV is
close to the lower bound of the interval which has been discussed
in the past.

We would like to thank all members of the KaoS Collaboration for
fruitful discussions especially A.~F\"orster, P.~Senger, C.~Sturm,
and F.~Uhlig.


\begin{thebibliography}{99}
\bibitem{SN} H.A. Bethe, Rev. Mod. Phys. {\bf 62} (1990) 801;
J.M. Lattimer, F.D. Swesty, Nucl. Phys. {\bf A 535} (1991) 331; H.
Shen, H. Toki, K. Oyamatsu, K. Sumiyoshi, Nucl. Phys. {\bf A 637}
(1998) 435.
\bibitem{you}D.H. Youngblood, H.L. Clark and Y.-W. Lui, Phys. Rev. Lett {\bf
84} (1999) 691.
\bibitem{mon}  J. Piekarewicz, Phys. Rev. {\bf C69} (2004) 041301
and references therein.
\bibitem{sto} H. St\"ocker and W. Greiner,
\newblock  Phys.~Reports~{\bf 137}, 278 (1986) and references therein.
\bibitem{aik} J. Aichelin and C.M. Ko, Phys. Rev. Lett. {\bf 55} (1985) 2661.
\bibitem{ant} A. Andronic et al., Phys. Lett. {\bf B612} (2005) 173.
\bibitem{paw2} P. Danielewicz, R. Lacey, W.G. Lynch, Science 298 (2002) 1592.
\bibitem{ai91} J.~Aichelin,
\newblock  Phys.~Reports~{\bf 202}, 233 (1991).
\bibitem{har} C. Hartnack et al., Eur. Phys. J. {\bf A1} (1998) 151.
\bibitem{crosss} E. Kolomeitsev et al., accepted J. of Phys. G,
nucl-th/0412037.
\bibitem{lu} C.L. Korpa and M.F.M. Lutz, submitted to Heavy Ion Physics,
nucl-th/0404088.
\bibitem{sturm} C. Sturm et al., (KaoS Collaboration), Phys. Rev. Lett. {\bf 86}
(2001) 39.
\bibitem{habil} C. Hartnack and J. Aichelin, Proc. Int. Workshop XXVIII
on Gross prop. of Nucl. and Nucl. Excit., Hirschegg, January 2000
edt. by M. Buballa, W. N\"orenberg, B. Sch\"afer and J. Wambach;
and  to be published in Phys. Rep.
\bibitem{fuchs}
C. Fuchs et al., Phys. Rev. Lett {\bf 86} (2001) 1794.
\bibitem{ko}
J.~Randrup and C.M.~Ko,
\newblock Nucl.~Phys. {\bf A 343}, 519 (1980).
\bibitem{jha} C. Hartnack, J. Aichelin, H. St\"ocker and W. Greiner
Phys. Rev . Lett. {\bf 72} (1994) 3767.
\bibitem{tsus} K. Tsushima et al., Phys. Lett. {\bf B337} (1994) 245;
Phys. Rev. {\bf C 59} (1999) 369. 
\bibitem{foer} A. F\"orster et al., (KaoS  Collaboration), Phys. Rev. Lett. {\bf
31} (2003) 152301; J. Phys. G. {\bf 30} (2004) 393; A.~F\"orster,
Ph.D.~thesis, Darmstadt University of Technology, 2003.
\bibitem{paw} G Stoicea et al., (FOPI Collaboration), Phys. Rev. Lett. {\bf 92} (2004)
072303.
\bibitem{and} A. Andronic et al., (FOPI Collaboration), Phys. Rev. {\bf C67} (2003)
034907.
\bibitem{hard}C. Hartnack et al., Nucl. Phys. {\bf A495} (1989)
303c.
\bibitem{hsqueeze} C. Hartnack et al., Mod Phys. Lett. A13 (1994)
1151.

\end{thebibliography}
\end{document}